# Raman and photoluminescence spectroscopic studies on structural disorder in oxygen deficient $Gd_2Ti_2O_{7-\delta}$ single crystals

M. Suganya[1,†], K. Ganesan[2,3,†,*], P.Vijayakumar[2], Amirdha Sher Gill[1,*], S.K.Srivastava[2], Ch. Kishan singh[2], R.M. Sarguna[2], P.K. Ajikumar[2], S.Ganesamoorthy[2,3]

[1]School of Science and Humanities, Sathyabama Institute of Science and Technology, Chennai, Tamil Nadu, India

[2]Materials Science Group, Indira Gandhi Centre for Atomic Research, Kalpakkam, Tamil Nadu, India

[3]Homi Bhabha National Institute, IGCAR, Kalpakkam, Tamil Nadu, India

[†] *These authors contributed equally*

[*]*Corresponding authors:* kganesan@igcar.gov.in *and* amirdhashergill@gmail.com

## Abstract

We report on Raman and photoluminescence spectroscopic studies on oxygen vacancy induced structural disorder in $Gd_2Ti_2O_{7-\delta}$ single crystals grown by optical floating zone technique under argon atmosphere. The oxygen vacancies in $Gd_2Ti_2O_{7-\delta}$ wafers decrease with thermal annealing in air atmosphere. The full width at half maximum of X-ray diffraction rocking curve decreases from 245 to 157 arc-second and the optical transmittance increases from 23 to 87 % (at 1000 nm) upon post growth thermal annealing. Raman spectroscopic studies reveal a monotonic increase in intensity of O-Gd-O ($E_g$) and Ti-O ($A_{1g}$) stretching modes with thermal annealing. Since these modes are associated with modulation of oxygen *x* parameter which is sensitive to Ti-O octahedron distortion, the increase in Raman intensity indicates an improvement in structural ordering of oxygen sub-lattice in $Gd_2Ti_2O_{7-\delta}$. Moreover, the photoluminescence studies also corroborate the Raman analysis in terms of reduction of structural defects associated with oxygen vacancies as a function of thermal annealing. This study demonstrates the effectiveness of using Raman spectroscopy to probe the structural disorder in $Gd_2Ti_2O_{7-\delta}$ crystals.

# 1. Introduction

The rare earth pyrochlore ($RE_2Ti_2O_7$) titanates are emerging as multi-functional materials because of their unique structural and intriguing physical properties which include magnetism, dielectric and ferroelectric phenomena, and also used as solid state electrolytes [1,2,11,3–10]. The gadolinium titanate ($Gd_2Ti_2O_7$, GTO) is one of the technologically important materials with diverse applications including photocatalysis, ionic conductivity, host material for optical emission and nuclear waste immobilization [4–9]. Since many of the physical and chemical properties of GTO emerge from the oxygen off-stoichiometry, the study on structure – property correlation is indispensable. In $Gd_2Ti_2O_7$ structure, there are two unique oxygen sites namely, 48$f$ and 8$b$. Six out of seven oxygen atoms of GTO reside at the position ($x,1/2,1/2$) in 48$f$ site which is coordinated with two $Gd^{3+}$ and two $Ti^{4+}$ cations. The remaining one oxygen is positioned at ($3/8,3/8,3/8$) in 8$b$ site and it is tetrahedrally coordinated with $Gd^{3+}$ cation . Further, the 8$a$ position is designated to the oxygen unoccupied interstitial site at ($1/8,1/8,1/8$).

The pyrochlore titanate ($RE_2Ti_2O_7$) structure is known to accommodate a large amount of lattice disorder because of its vacant 8$a$ site and the complex chemical structure with alternating strong (Ti-O) and weak (RE-O) bondings. Further, the oxygen off-stoichiometry is one of the common issues in pyrochlore materials due to the high temperature growth process. A small deviation in oxygen stoichiometry in pyrochlore structure could lead to the formation of cation antisite disorder (6.7 %), vacancies mainly at 48$f$(O) (15%) and 8$b$ (9 %) sites and two new anion defect sites (O3, O4), as reported for $Lu_2Ti_2O_{6.43}$ using neutron diffraction [11]. It is known that the oxygen vacancy defects and cationic disorder in $Gd_2Ti_2O_7$ can enhance ionic conductivity, magnetic behaviour and photocatalytic activity significantly [4,6]. Thus, the oxygen vacancies and other defects would provide an additional opportunity to tune and control the desired structural and physical properties of pyrochlore structures. Such structural defects can be easily created in crystals by growing in oxygen deficient environment or annealing at high temperature under reducing atmosphere [8,11]. Hence, the evaluation of oxygen vacancies in pyrochlore structures becomes essential in order to have a meaningful interpretation of physical properties from different set of samples. However, the evaluation of oxygen vacancies and structural disorder in pyrochlore structures are not an easy task due to the complexity of its structure.

The pyrochlore oxides have a complex structure and the complexity increases with the introduction of oxygen vacancies. Solving the structure of pyrochlore oxides is non-trivial and is studied successfully by neutron diffraction [11–13]. X-ray diffraction and Raman spectroscopy are also commonly used to probe the disorder induced structural properties of pyrochlore structures. Mostly, the studies are reported on the defect induced order – disorder transition in rare earth or transition metal doped pyrochlore / fluorite structures [11,14,15]. Also, these studies are extended to the nanostructured pyrochlores prepared and processed at different temperatures [9,16]. In addition, there are also several intensive studies available on the relation between oxygen off-stoichiometry and physical properties of $Gd_2Ti_2O_7$ or other RE titanate structures [4,5,9,11,17,18]. Despite these extensive studies on pyrochlore oxides, the effect of structural defects on physical properties are not well understood and it could be due to lack of understanding in oxygen off-stoichiometry. Chen et al [19] had reported the occurrence of simultaneous cationic antisite disorder and anion disorder in $Gd_2(Ti_{1-x}Zr_x)_2O_7$ by probing the O 1$s$ state using X-ray photoelectron spectroscopy. They reported the two O 1$s$ peaks corresponding to 48$f$ and 8$b$ sites merge together with increase in disorder by $Zr$ doping. Similarly, in principle, Raman spectroscopy should also be able to probe anion disorder with ease in pyrochlore structures. Further, Raman spectroscopy is a powerful, simple and non-destructive technique to study the local symmetry and structural disorder in materials. However, a systematic study on anion disorder related to oxygen vacancies in $Gd_2Ti_2O_{7-\delta}$ or any other pyrochlore oxides is not reported so far using Raman spectroscopy.

In this report, we investigate the effect of oxygen off-stoichiometry in $Gd_2Ti_2O_{7-\delta}$ crystals by optical absorption, Raman and photoluminescence spectroscopy. To prepare $Gd_2Ti_2O_{7-\delta}$ crystals with different oxygen vacancies, the crystal was grown under argon atmosphere that would lead to high oxygen vacancies in the structure. Subsequent thermal annealing of the crystals at 1000 ºC for 20, 30 and 50 hrs under oxygen containing atmosphere had resulted in obtaining $Gd_2Ti_2O_{7-\delta}$ with different oxygen concentrations. The oxygen stoichiometry of the as-grown and 50 h annealed wafers were quantified using thermogravimetric analyser (TGA). Further, a systematic study on the structural ordering, within the pyrochlore structure, was evaluated by Raman spectroscopy and the results were further corroborated with TGA, rocking curve analysis, optical absorption and photoluminescence (PL) studies.

## 2 Experimental methods

Single crystals of $Gd_2Ti_2O_{7-\delta}$ were grown by optical floating zone technique in Ar atmosphere and the details were discussed elsewhere [20,21]. First, the polycrystalline powders were prepared by standard solid state reaction at 1300 °C with several intermediate grinding using stoichiometric oxides, $Gd_2O_3$ and $TiO_2$ of 4N purity. After confirming the phase formation, single crystals were grown in Ar atmosphere at optimized growth parameters using the synthesized polycrystalline materials. The feed / seed rods were counter rotated at ~ 30/30 rpm and the growth rate was varied in the range 4-8 mm/h. A crack-free crystal with uniform diameter could be grown in Ar atmosphere. The grown crystals were completely opaque due to oxygen off-stochiometry. The cut and polished wafers were further annealed at 1000 °C for 20, 30 and 50 h in air atmosphere, in order to obtain $Gd_2Ti_2O_{7-\delta}$ crystal with different oxygen vacancies.

Powder X-ray diffraction (PXRD) studies were carried out using STOE-XRD instrument and rocking curve measurements were performed in Brucker D8 Discover with CuKα source. Oxygen stoichiometry of the $Gd_2Ti_2O_{7-\delta}$ crystal was measured for as-grown and 50 h annealed wafers by TGA using Setaram SETSYS 16/18. Fine powders of $Gd_2Ti_2O_{7-\delta}$ wafer of about 30 mg was loaded into platinum crucible and heated upto 500 °C at a heating rate of 1.2 °C/min under air atmosphere. The initial oxygen stoichiometry of the compound was estimated from the change in weight gain from TGA data. Here, the final saturated mass was considered as maximum oxidation state of the compound. Raman spectra were recorded using micro-Raman spectrometer (InVia, Renishaw, UK) with 2400 grooves / mm grating at 100X objective magnification and laser excitation wavelength of 532 nm in back scattering geometry. The laser power was kept below 1 mW to avoid the laser induced heating of samples. The acquired spectra were analyzed using Wire 4.2 software. The UV-Vis-NIR studies were performed using Lambda 35 Perkin Elmer in the wavelength range of 200-1100 nm.

## 3 Results
### 3.1. X-ray diffraction analysis

The PXRD patterns of as-synthesized compound by solid state reaction and powdered 50 h annealed $Gd_2Ti_2O_7$ wafer are shown in Fig. 1a. These two diffraction patterns match well with reported JCPDS card no. 00-023-0259 which confirms the $Gd_2Ti_2O_7$ phase formation. The X-ray diffraction rocking curves measured on the as-grown and annealed wafers are shown in Fig. 1b. The full width at half maximum (FWHM) of rocking curve decreases systematically from 245 to 157 arcsec with annealing duration as shown in inset of Fig. 1b. This behavior clearly indicates the improvement of structural quality of the crystals with thermal annealing. Further, the sharp rocking curves without any shoulder reveal the absence of sub-grain boundaries in the grown $Gd_2Ti_2O_{7-\delta}$ crystals.

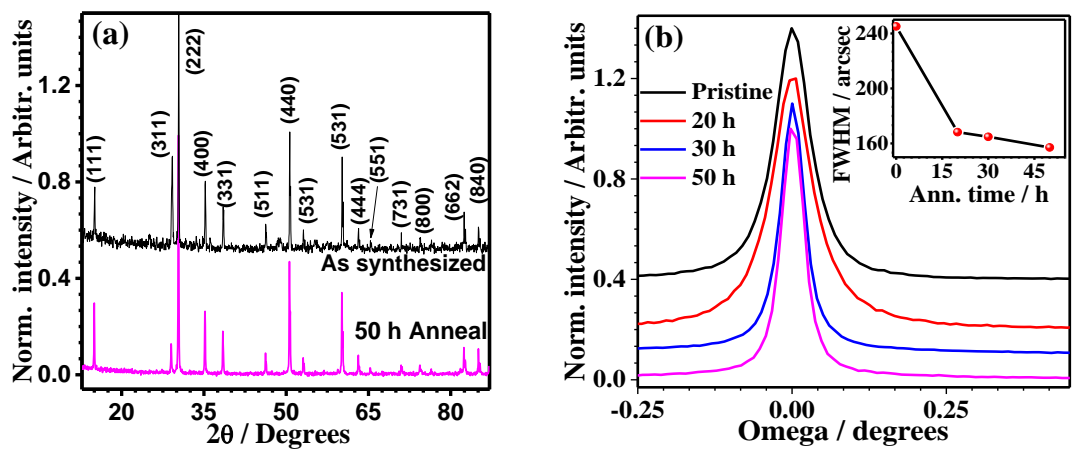

Fig. 1. (a) Powder X-ray diffraction pattern of as synthesized compound and powdered wafer of 50 h annealed $Gd_2Ti_2O_{7-\delta}$. (b) X-ray diffraction rocking curves of as-grown and annealed wafers of $Gd_2Ti_2O_{7-\delta}$ for different durations. The inset shows the variation of FWHM as a function of annealing time of $Gd_2Ti_2O_{7-\delta}$ wafers.

## 2. Thermogravimetric analysis

Figure 2 depicts the TGA data for the powdered wafer of pristine and 50 h annealed $Gd_2Ti_2O_{7-\delta}$. The pristine sample loses weight of 0.2 mg initially while heating upto ~ 240 °C. The subsequent heating enables the sample to gain additional mass of ~ 0.22 mg and it reaches a saturation value above the temperature of ~ 330 °C. On the other hand, the 50 h annealed wafer displays a continuous gain of mass from ~ 150 °C and reaches a saturated value above 300 °C with a net weight gain of ~ 0.04 mg. Here, both the samples exhibit a mass gain in single step upto temperature of ~ 330 °C. This mass gain is attributed to the oxidation of $Ti^{3+}$ into $Ti^{4+}$ state by oxygen diffusion in the $Gd_2Ti_2O_{7-\delta}$ lattice. Further, the mass gain is consistent with oxidation of $Ti^{3+}$ for a similar pyrochlore titanate structures which shows a weight gain in the temperature range of 200 to 300 °C [22]. Further, the oxygen vacancy ($\delta$) in $Gd_2Ti_2O_{7-\delta}$ is calculated based on the mass gain. The molar ratio of the oxygen vacancy is found to be ~ 0.23 and 0.04 for pristine and 50 h annealed wafers, respectively. Here, the weight gain from temperature 240 to 330 °C is used for calculating $\delta$ for pristine wafer. The reduction of weight in pristine wafer is attributed to the removal of physisorbed environmental molecules during heating up to 240 °C. This occurs in pristine wafer because it has large oxygen vacancies that could easily adsorb moisture. On the other hand, the 50 h annealed wafer does not have much oxygen vacancies and hence, there is no loss of weight during initial heating. Thus, the TGA study clearly reveals the reduction of oxygen vacancies after thermal annealing.

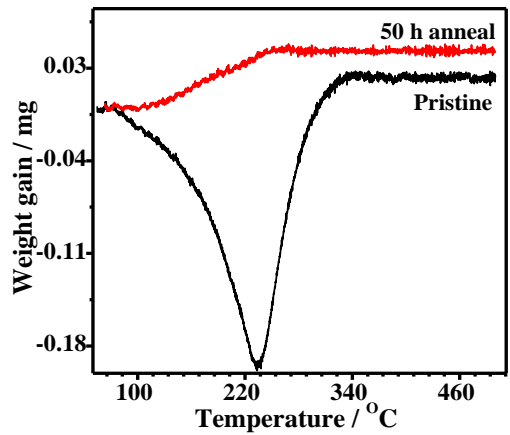

Fig. 2. The change of weight as a function of temperature of $Gd_2Ti_2O_{7-\delta}$ powders of as-grown and 50 h annealed wafers.

### 3.2. Raman spectroscopy

Figure 3 depicts the Raman spectra of pristine and annealed $Gd_2Ti_2O_{7-\delta}$ wafers and the spectra appear nearly the same. The typical Raman spectrum shows six Raman active modes of pyrochlore structure and are labeled as M1 (203 cm$^{-1}$, $F_{2g}$), M2 (311 cm$^{-1}$, $F_{2g}$), M3 (328 cm$^{-1}$, $E_g$), M4 (450 cm$^{-1}$, $F_{2g}$), M5 (517 cm$^{-1}$, $A_{1g}$), and M6 (540 cm$^{-1}$, $F_{2g}$) [15,23,24]. The Raman spectra also display high wavenumber modes M7, M8 and M9 modes at ~ 673, 696 and 860 cm$^{-1}$ which are attributed to second order Raman scattering [23]. Note that the absolute Raman intensity of the annealed $Gd_2Ti_2O_{7-\delta}$ crystals is slightly higher than that of as-grown crystal. However, the intensity is normalized to high intensity M2 mode for comparison. Further, the Raman spectra of $Gd_2Ti_2O_{7-\delta}$ wafers are deconvoluted using wire 4.2 software with mixed Lorentzian and Gaussian line profiles in the wavenumber range of 170 – 380 and 480 – 620 cm$^{-1}$ for detailed analysis. The best fit Raman parameters for M1-M3, M5 – M7 modes are shown in Table 1. Figure 4 shows the intensity ratios (peak height and area under the curve ) of M3 and M5 modes with M2 mode which increase as a function of annealing duration. The increase in Raman intensity with annealing duration is attributed to the decrease of distortion in anion sub-lattice of $Gd_2Ti_2O_{7-\delta}$ by oxidation.

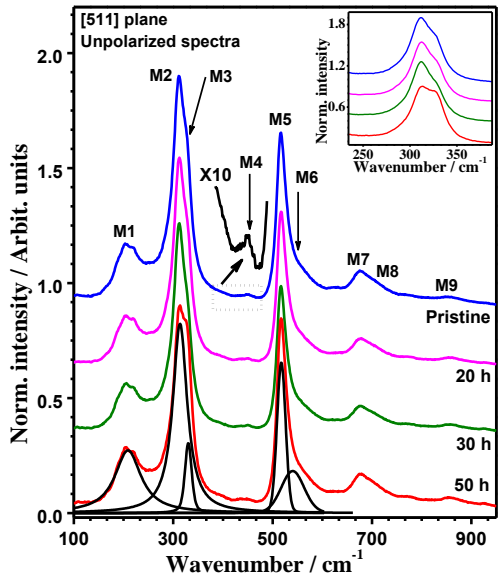

Fig. 3. Raman spectra of as-grown and thermal annealed $Gd_2Ti_2O_{7-\delta}$ wafers. The intensity of these spectra are normalized with 311 cm$^{-1}$ mode and stacked vertically for clarity. A magnified part of the M4 mode is shown in the graph with arrow mark. The inset plot displays the magnified part of the Raman spectra in the wavenumber range of 250 – 380 cm$^{-1}$. The de-convoluted peaks are shown for 50 h annealed $Gd_2Ti_2O_{7-\delta}$ wafer.

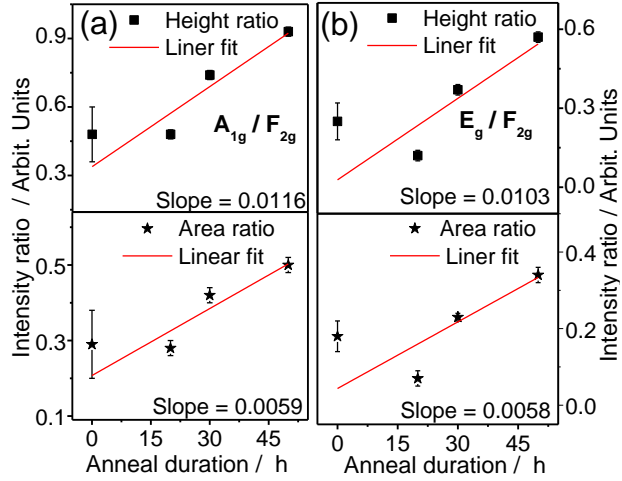

Fig. 4. The peak height and area intensity ratios of Raman modes, (a) $A_{1g}$ (517 cm$^{-1}$) and (b) $E_g$ (328 cm$^{-1}$) with respect to $F_{2g}$ mode ( 311 cm$^{-1}$ ) as a function of annealing duration. The straight line in the plots represents the linear fit to the data.

Table 1. Raman fit parameters of $F_{2g}$, $E_g$ and $A_{1g}$ modes of $Gd_2Ti_2O_{7-\delta}$ wafers with different oxygen stoichiometry.

| Sample | M1, $F_{2g}$ [cm$^{-1}$] | | M2, $F_{2g}$ [cm$^{-1}$] | | M3, $E_g$ [cm$^{-1}$] | | M5, $A_{1g}$ [cm$^{-1}$] | | M6, $F_{2g}$ [cm$^{-1}$] | | M7, $F_{2g}$ [cm$^{-1}$] | |
|---|---|---|---|---|---|---|---|---|---|---|---|---|
| | Position | FWHM | Position | FWHM | Position | FWHM | Position | FWHM | Position | FWHM | Position | FWHM |
| Pristine | 206.0±0.5 | 42.2±2.0 | 311.4±0.1 | 26.6±0.2 | 328.6±0.1 | 19.1±0.3 | 516.5±0.1 | 17.6±0.3 | 528.2±0.2 | 36.8±1.5 | 673.4±0.3 | 28.5±1.5 |
| 20 h | 204.7±0.6 | 42.2±0.7 | 310.6±0.1 | 25.8±0.1 | 328.1±0.3 | 18.2±1.2 | 515.5±0.1 | 17.1±0.2 | 527.5±0.1 | 37.1±1.2 | 672.9±0.4 | 30.4±0.2 |
| 30 h | 205.0±0.2 | 44.0±0.6 | 311.3±0.1 | 26.4±0.1 | 328.6±0.1 | 18.7±0.2 | 515.8±0.1 | 16.7±0.2 | 527.4±0.3 | 33.6±1.3 | 673.2±0.2 | 27.6±1.5 |
| 50 h | 204.1±0.5 | 44.8±2.0 | 311.2±0.1 | 29.0±0.2 | 328.2±0.1 | 19.1±0.1 | 515.5±0.1 | 17.2±0.3 | 527.4±0.4 | 38.1±2.9 | 673.1±0.3 | 31.7±1.2 |

### 3.3. UV-Vis-NIR spectroscopy

Figure 5 shows the optical transmittance spectra of as-grown and annealed wafers of thickness ~1 mm. The transmittance increases drastically from 23 to 87 % at 1000 nm for annealed wafers. Also, the annealed wafers have high transmittance in the wavelength range of 370 – 1200 nm. However, the as-grown wafer exhibits a strong absorption in the wavelength range 400 – 900 nm as can be evidenced from inset of Fig. 5. This strong absorption is correlated to intervalance transition to $Ti^{3+}$ charge state which occurs due to the oxygen vacancies and associated cationic anti-site disorder. Upon thermal annealing, $Ti^{3+}$ ions oxidize into $Ti^{4+}$ state and hence, the optical transmittance increases with increase in structural ordering in $Gd_2Ti_2O_{7-\delta}$ lattice. Also, Fermi level shifts towards conduction band minimum as a function of oxygen vacancy concentration [18]. The estimated band gap is found to be about 3.5 eV for pristine and a constant of 3.6 eV for all annealed $Gd_2Ti_2O_{7-\delta}$ wafers and these values are consistent with reported data [6]. A slight lower bandgap of as-grown wafer is attributed to the $Ti^{3+}$ ions induced localized states within the bandgap. Also, the transition edge near the bandgap becomes steeper for annealed wafers as compared to the as-grown one, indicating the improvement of structural and optical quality of the crystals with annealing time.

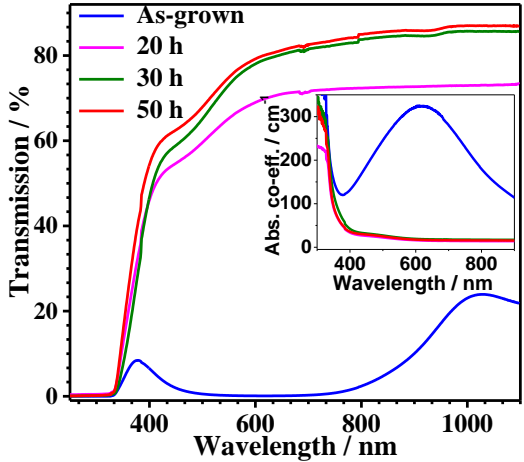

Fig. 5. UV-Vis-NIR transmission spectra of as-grown and annealed wafers of $Gd_2Ti_2O_{7-\delta}$. The inset shows the variation of absorption co-efficient of $Gd_2Ti_2O_{7-\delta}$ wafers annealed for different durations, as calculated from the transmission spectra.

### 3.4. Photoluminescence spectroscopy

Figure 6a displays the PL emission spectra of the as-grown and thermally annealed wafers of $Gd_2Ti_2O_{7-\delta}$ for different durations. The spectra are recorded under identical experimental conditions with excitation wavelength of 300 nm. The as-grown wafers display a very weak and broad emission band at ~ 400 nm and the PL emission intensity is almost zero for the wavelength above 580 nm (curve 1 in Fig 6a) due to high oxygen vacancies, $Ti^{3+}$ ionic states and other point defects. These defects create large density energy levels within the forbidden bandgap and they undergo non-radiative optical transition. Hence, PL emission is negligibly small. This observation is also supported by the strong absorption of light in the 400 – 900 nm as discussed in UV-Vis spectroscopy. Upon annealing for 20 hrs, a significant amount of oxygen vacancies and other point defects decrease in the lattice. Consequently, the density of defect energy levels within the bandgap decreases enormously. But, still a large defect states are present in the forbidden bandgap and they undergo radiative transition when excited with high energy photons. These radiative transition induces very broad PL emission in the wavelength range of 400 – 700 nm as shown in curve 2 of Fig. 6a. Upon further annealing for 30 and 50 hrs, these defect induced energy levels decrease substantially leading to minimal radiative transition between defect levels and valance / conduction band. Hence, the broad background emission in the wavelength range of 400 – 700 nm decreases with annealing duration, as shown in curves 3 and 4 of Fig. 6a. Also, two broad PL emission bands are observed about 412 and 438 nm for 30 and 50 h annealed wafers. These emission bands may be attributed to the oxygen related F and

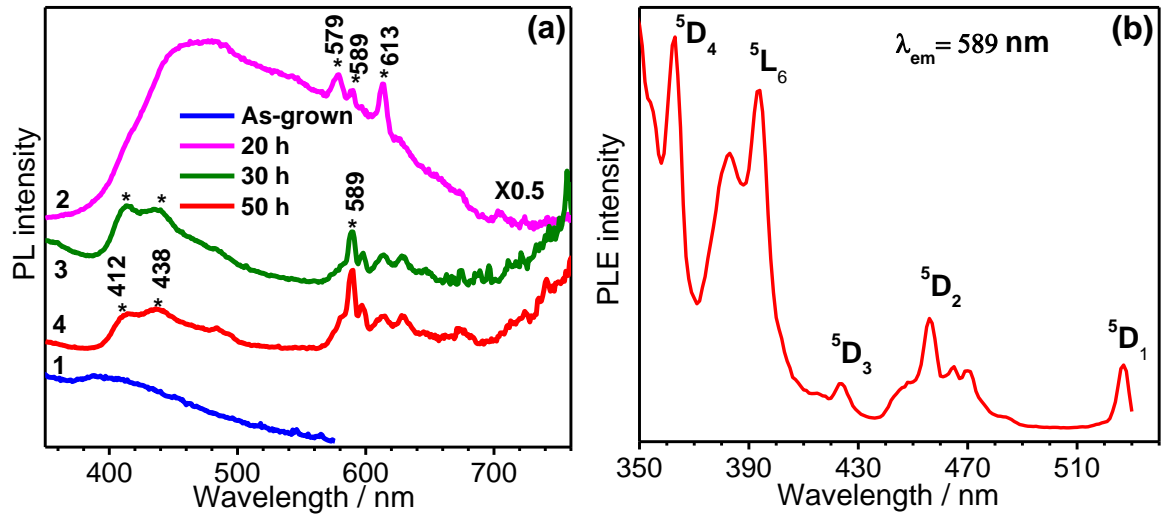

Fig. 6 (a) Photoluminescence (PL) emission spectra of as-grown and thermally annealed wafers of $Gd_2Ti_2O_{7-\delta}$. The curve nos. 1, 2, 3 and 4 in the graph (Fig. 6a) represent the wafer that is annealed for 0, 20, 30 and 50 h, respectively. The spectra are stacked vertically for sake of clarity without subtracting background. The PL spectrum of 20 h annealed wafer is multiplied by 0.5 to bring it to the scale for comparison. (b) Typical PL excitation spectrum of 50 h annealed $Gd_2Ti_2O_{7-\delta}$ wafer for emission of 589 nm.

$F^+$ color centers, respectively. Further, these emission bands are not observed on as-grown and 20 h annealed $Gd_2Ti_2O_{7-\delta}$ wafers since these wafers consist of large disorder in them. Apart from the defect related emission bands, we also observe sharp emission lines at 579 and 613 nm for 20 h annealed wafers and a sharp emission at 589 nm along with several low intensity bands for both 30 and 50 h annealed wafers. In order to study the origin of these sharp emission lines, PL excitation (PLE) spectra are recorded for the annealed $Gd_2Ti_2O_{7-\delta}$ wafers. Fig. 6b displays a typical PLE spectrum for the emission wavelength of 589 nm and it exhibits several characteristics emission lines at 363, 424, 456 and 527 nm corresponding to the quantum states of $^5D_4$, $^5D_3$, $^5D_2$ and $^5D_1$, respectively and these states indicate the presence of unintentional dopant $Eu^{3+}$ ions in the lattice [7,9,25]. Thus, the PL emission lines at 579, 589 and 613 nm are assigned to the spin forbidden *f-f* transitions associated with $^5D_0 \rightarrow {^7F_J}$ (J=0,1 and 2) states, respectively.

## 4. Discussion

In pyrochlore structures, the Raman modes only arise due to the vibrations of oxygen at *48f* (O) and *8b* (O′) positions since Gd and Ti cations do not contribute to Raman active modes due to their centro-symmetric positions. In $Gd_2Ti_2O_{7-\delta}$, the most intense peak at 311 cm$^{-1}$ (combined M2 and M3 modes) originate from O′-Gd-O′ bending vibrations in $GdO_8$ polyhedron. The characteristic $E_g$ and $A_{1g}$ (M3 & M5) modes are particularly interesting in pyrochlore structures since they are associated with the modulation of oxygen *x* parameter which is connected with Ti-O coordinated bonds in $TiO_6$ octahedron distortion at *48f* (O). Further, M5 mode can be regarded as symmetric breathing motion of the oxygen at 48*f*(O) octahedron towards the vacant 8*a* site, i.e. the stretching of 48*f* O−vacancy bond at 8*a* site. Moreover, the $E_g$ mode also involves with asymmetric stretching of *48f*(O) octahedron centered at vacant 8*a*(O) site [23]. The intensity of M4 mode (450 cm$^{-1}$) is very weak and a magnified part of the spectrum is indicated by an arrow mark in the inset of Fig. 3.

It is well known that the Raman intensity of a particular mode is directly proportional to the local polarizability of the molecule and the concentration of the active molecule. Further, the polarizability of the molecule decreases with increase in electron density of the molecule. Here, the as-grown $Gd_2Ti_2O_{7-\delta}$ wafer is found to have oxygen vacancies of $\delta = \sim 0.23$. Consequently, two O atoms are vacant out of six O atoms at 48$f$(O) position in 15 % TiO$_6$ octahedron. Further with additional O3 oxygen defect in anion sub-lattice, TiO$_6$ octahedron structure becomes five coordinated distorted trigonal-bipyramidal structure. Moreover, the oxygen vacancies transforms the cation from Ti$^{4+}$ into Ti$^{3+}$ charge state, as reported earlier in literature on the structure of oxygen deficient Lu$_2$Ti$_2$O$_{6.43}$ by Blundred et al [11] using neutron diffraction. This structural distortion in oxygen sub-lattice induces change in local symmetry with shorter 48f(O) – O3 bonds and increased electron density from Ti$^{3+}$ ionic states. This change in local symmetry decreases the polarizability of the molecule. Hence, Raman intensity of the A$_{1g}$ mode is lowest for as-grown Gd$_2$Ti$_2$O$_{7-\delta}$ wafers. Upon thermal annealing, a considerable amount of oxygen diffuses into the lattice. This results in the reduction of vacancy induced distortion at Ti-O anion sub-lattice leading to a significant increase in polarizability which would eventually result in increase of Raman intensity of A$_{1g}$ mode. Similarly, the intensity of E$_g$ mode also increases with improvement in the structural ordering of anion sub-lattice at GdO$_8$ polyhedron with thermal annealing. Also, it has been reported in literature that the mixed pyrochlore structure with oxygen vacancies displays high dielectric loss which is correlated with the decrease in polarizability of the molecule and it supports our Raman observation [26,27]. We note here that, for the first time, a systematic increase in Raman intensity of A$_{1g}$ and E$_g$ modes is demonstrated with annealing time indicating the structural ordering of anion sub-lattice in Gd$_2$Ti$_2$O$_{7-\delta}$.

As shown in Table 1, the peak position and FWHM of the Raman modes do not vary significantly for Gd$_2$Ti$_2$O$_{7-\delta}$ crystals annealed for different durations. On the other hand, the optical property of the crystals varies significantly with thermal annealing as evidenced by UV-vis-NIR absorption spectroscopy. Note that UV-Vis absorption / transmission spectrum is based on electronic transitions between available energy levels which are very sensitive to defects that arise due to oxygen off-stoichiometry. On the other hand, the phonon modes may not reflect a similar change with respect to oxygen off-stoichiometry. However, the intensity of selected modes gradually increases as a function of thermal annealing as shown in Fig. 4. These observations reveal that the structural quality of the crystals are excellent even in the as-grown crystals with oxygen vacancies. Though the thermal annealing process improves the oxygen stoichiometry, the phonon modes are not very sensitive as similar to optical absorption with respect to oxygen off-stoichiometry.

Moreover, a careful observation on PL emission spectra (Fig. 6a) reveal that the electric – dipole induced transitions ($^5D_0 \rightarrow {}^7F_0$ and $^5D_0 \rightarrow {}^7F_2$) at 579 and 613 nm emission are more dominant than 589 nm emission for 20 h annealed wafer. On the other hand, 589 nm emission line is more dominant over 579 and 613 nm lines for 30 and 50 h annealed wafers. Further, the 50 h wafer displays higher intensity for 589 nm emission line and lower intensity for 613 nm emission line as compared to 30 h annealed wafer. We note here that the doublet emission at 589 nm occurs due to magnetic – dipole transition ($^5D_0 \rightarrow {}^7F_1$ transition). According to the Laporte rule, only the $^5D_0 \rightarrow {}^7F_1$ transition is allowed when the $Eu^{3+}$ is situated at centro-symmetric site of the lattice [25]. However, the forbidden electric – dipole transitions associated with 579 and 613 nm emission are activated when the crystal consists of considerable amount of defects that can lead to the distortion at $Eu^{3+}$ sites. Here, $Eu^{3+}$ ion substitutes $Gd^{3+}$ ion since the ionic radii of $Eu^{3+}$ is comparable to $Gd^{3+}$, and much larger than $Ti^{4+}$ [ionic radii, R($Eu^{3+}$, CN=8) = 1.066 Å, R($Gd^{3+}$, CN=8) = 1.053 Å, and R(Ti4+, CN=6) = 0.605 Å] [28] and $Gd^{3+}$ site is also coordinated with oxygen at 48f site at $GdO_8$ polyhedron. Hence, the oxygen vacancy induced structural distortion at $Eu^{3+}$ sites invokes the spectroscopically forbidden electric – dipole transitions corresponding to the emissions at 579 and 613 nm. Upon thermal annealing for 30 and 50 h, the structural distortion at *Gd* cation site decreases significantly and hence, the intensity of the forbidden emission bands (579 and 613 nm) decreases and also, the intensity of spectroscopically allowed magnetic – dipole transition band (589 nm) increases systematically. The source of unintentional Eu dopants would be from the raw material, $Gd_2O_3$. Since we used $Gd_2O_3$ with 99.99 % of purity, the maximum possible *Eu* concentration in $Gd_2Ti_2O_7$ is about 100 ppm. Due to this low doping concentration, *Eu* is not detectable by X-ray diffraction and Raman spectroscopy. However, the *Eu* impurities with < 100 ppm create characteristic PL emission lines in $Gd_2Ti_2O_7$ crystal. Though the $Eu^{3+}$ ions are unintentional impurity in the $Gd_2Ti_2O_{7-\delta}$ crystal, it serves perfectly to monitor the structural distortion in the lattice by characteristic PL emission bands. Thus, the PL spectroscopy corroborates Raman spectroscopy with direct evidence for the structural ordering in the oxygen sub-lattice.

## 4. Conclusions

Gd$_2$Ti$_2$O$_{7-\delta}$ single crystals with oxygen off-stoichiometry were successfully grown in Ar atmosphere using optical floating zone technique. Thermal annealing decreases the oxygen vacancies and consequently, the structural and optical quality of the crystals improve significantly. The most striking observation of this study is that the Raman intensity of E$_g$ and A$_{1g}$ modes, which are associated with the oxygen vacancies at *48f* (O) position, increases monotonically with annealing duration. This observation reveals that the vacancy induced distortion in oxygen sub-lattice of Gd$_2$Ti$_2$O$_{7-\delta}$ decreases systematically with thermal annealing. In addition, the photoluminescence studies also support the improvement in the structural ordering in Gd$_2$Ti$_2$O$_{7-\delta}$ crystals by monitoring characteristic emission bands from the unintentional Eu$^{3+}$ impurities. Even though thermal annealing is known to improve the structural quality of crystal, for the first time, Raman and photoluminescence spectroscopies are used to demonstrate the anion vacancy induced distortion in oxygen sub-lattice of pyrochlore structures.

## Acknowledgement

The authors M.S and A.S.G acknowledge University Grants Commission (UGC), India under UGC-DAE-CSR (CSR-KN/CRS-87/2016-17/1128) for financial support and Dr.N.V. Chandrasekar, Scientist in-charge, UGC-DAE-CSR, Kalpakkam node for encouragement. Also, authors thank Mrs. Sunitha Rajakumari, MSG/IGCAR for crystal polishing & rocking curve measurements. One of the authors, K.G, thank Dr. G. Mangamma and Dr. S. Dhara SND/MSG/IGCAR for their support and encouragement.